\documentclass[twocolumn, 9pt]{article}

\usepackage[margin=0.75in, columnsep=0.25in]{geometry}
\usepackage{times}

\usepackage{graphicx}
\usepackage{caption}
\usepackage{subcaption}
\usepackage{booktabs}
\usepackage{makecell}
\usepackage{multirow}
\usepackage{tabularx}
\usepackage{array}
\usepackage{ragged2e}

\usepackage{amsmath, amssymb, amsthm}
\usepackage{bm}
\usepackage{algorithm}
\usepackage{algpseudocode}

\usepackage{enumitem}
\usepackage{titlesec}

\titleformat{\section}
{\large\bfseries}
{\thesection}
{0.5em}
{}
\titleformat{\subsection}
{\normalsize\bfseries}
{\thesubsection}
{0.5em}
{}
\titleformat{\subsubsection}
{\normalsize\itshape}
{\thesubsubsection}
{0.5em}
{}
\titlespacing*{\section}{0pt}{2.5ex}{1.5ex}
\titlespacing*{\subsection}{0pt}{2ex}{1ex}
\titlespacing*{\subsubsection}{0pt}{1.5ex}{0.5ex}

\usepackage{cite}
\usepackage{url}

\usepackage{hyperref}
\hypersetup{
	colorlinks=true,
	linkcolor=blue,
	filecolor=magenta,
	urlcolor=cyan,
	citecolor=blue,
}

\usepackage[capitalize,noabbrev]{cleveref}

\usepackage{authblk}

\newcolumntype{C}[1]{>{\centering\arraybackslash}m{#1}}

\title{\textbf{Crypto-ncRNA: a bio-inspired post-quantum cryptographic primitive exploiting RNA folding complexity}}

\author[2]{Xu Wang\thanks{These authors contributed equally to this work.}}
\author[1,3]{Yiquan Wang\protect\footnotemark[1]}
\author[4]{Tin-Yeh Huang\protect\footnotemark[1]}
\author[5]{Zhaorui Jiang}
\author[1]{Kai Wei\thanks{Corresponding author: \href{mailto:kaiwei@xju.edu.cn}{kaiwei@xju.edu.cn}}}

\affil[1]{Xinjiang Key Laboratory of Biological Resources and Genetic Engineering, College of Life Science and Technology, Xinjiang University, Urumqi, Xinjiang, China}
\affil[2]{Tsinghua University-Peking University Joint Center for Life Sciences, Tsinghua University, Beijing, China}
\affil[3]{College of Mathematics and System Science, Xinjiang University, Urumqi, Xinjiang, China}
\affil[4]{Department of Industrial and Systems Engineering, Faculty of Engineering, The Hong Kong Polytechnic University, Hong Kong SAR, China}
\affil[5]{School of Environment and Energy, Shenzhen Graduate School, Peking University, Shenzhen, China}

\date{}

\begin{document}
	
	\maketitle
	
	\begin{abstract}
		The imminent realization of fault-tolerant quantum computing precipitates a systemic collapse of classical public-key infrastructure and necessitates an urgent transition to post-quantum cryptography. However, current standardization efforts predominantly rely on structured mathematical problems that may remain vulnerable to unforeseen algorithmic breakthroughs, highlighting a critical need for fundamentally orthogonal security paradigms. Here, we introduce \emph{Crypto-ncRNA} as a biophysically inspired cryptographic primitive that exploits the thermodynamic complexity of non-coding RNA folding as a computational work-factor amplifier. By leveraging the rugged energy landscape inherent to RNA secondary structure prediction, a problem intractable to rapid inversion, we establish a security foundation independent of conventional number-theoretic assumptions. We validate this approach by mapping the folding problem to a Quadratic Unconstrained Binary Optimization model and demonstrate theoretical resilience against quantum optimization attacks including the Quantum Approximate Optimization Algorithm. Functioning as a symmetric key encapsulation and derivation primitive dependent on pre-shared seeds, Crypto-ncRNA achieves throughputs competitive with software-based Advanced Encryption Standard implementations. By utilizing the generated high-entropy keys within a standard stream cipher framework, it exhibits ciphertext entropy that satisfies rigorous NIST SP 800-22 statistical standards. These findings not only articulate a novel bio-computational pathway for cryptographic defense but also provide a rigorous algorithmic blueprint for future physical realization, demonstrating that the thermodynamic complexity of biological systems offers a robust and physically grounded frontier for securing digital infrastructure in the post-quantum era.
	\end{abstract}
	
	\vspace{0.5em}
	
	\section{Introduction}
	The imminent realization of fault-tolerant quantum computing precipitates a systemic collapse of classical public-key infrastructure and necessitates an urgent transition to post-quantum cryptography. The security of ubiquitous protocols relies predominantly on the computational difficulty of integer factorization, a problem that becomes efficiently solvable via Shor's algorithm \cite{Shor1994}. This systemic vulnerability has catalyzed global standardization efforts to identify cryptographic solutions capable of resisting quantum adversaries \cite{chen2016report, alagic2022status, alagic2019status, Balamurugan2021, bavdekar2023post, kumar2020post}. The urgency of this transition is further amplified by the prospective threat wherein adversaries stockpile encrypted data for future decryption \cite{ott2019identifying, singh_managing}. While current standardization initiatives focus on mathematical constructions such as lattice-based and code-based cryptography \cite{Alagic2020, Bernstein2017, basu2019nist}, these approaches typically rely on structured algebraic hardness assumptions. To hedge against unforeseen algorithmic breakthroughs that might compromise these mathematical foundations, there is a critical imperative to explore fundamentally orthogonal security paradigms that derive their robustness from physical rather than purely number-theoretic complexities.
	
	The intersection of molecular biology and computational theory provides a fertile ground for such alternative paradigms. Since the seminal demonstration of molecular computation for solving combinatorial problems \cite{adleman1994molecular}, biomolecules have been recognized not merely as genetic storage media but as physical substrates capable of massive parallel processing. Among these biological processes, the folding of ribonucleic acid represents a problem of profound computational intricacy. Unlike the idealized folding funnels often observed in protein dynamics which guide the molecule efficiently toward a native state \cite{Leopold1992, Bryngelson1995, Onuchic1997, Kuzu2025}, RNA folding landscapes are frequently characterized by severe ruggedness and deep kinetic traps \cite{Solomatin2010}. The prediction of the Minimum Free Energy secondary structure for an RNA sequence involves navigating this complex thermodynamic terrain. While dynamic programming algorithms have been established to solve standard folding problems \cite{Zuker1981, Zuker1984}, the inclusion of complex topological features such as pseudoknots dramatically escalates the computational demand, rendering general structure prediction and its inverse problem mathematically intractable \cite{Rivas1999, Akutsu2000, Masuki2026}.
	
	Here, we introduce \emph{Crypto-ncRNA}, a biophysically inspired cryptographic primitive that utilizes thermodynamic complexity as a computational work-factor amplifier. Distinct from lattice-based or code-based constructs, which rest upon structured algebraic hardness that remains theoretically vulnerable to hidden symmetries or algorithmic inversion, the security of Crypto-ncRNA is predicated on the stochastic and chaotic nature of high-dimensional energy landscapes. This introduces a source of entropy that is fundamentally orthogonal to traditional number-theoretic assumptions, thereby providing an irreplaceable layer of defense against mathematical cryptanalysis. By incorporating structural dependencies into the encryption schema, we transform the decryption process into a validation task that necessitates traversing the rugged energy landscape of a simulated RNA molecule. This design effectively treats the biophysical simulation as a computational barrier analogous to a one-way function \cite{katz2007introduction}. To verify a potential key, an adversary is forced to execute a computationally expensive simulation, imposing an $O(N^3)$ complexity penalty characteristic of algorithms like LinearFold \cite{huang2019linearfold}. We validate this approach by mapping the folding problem to a Quadratic Unconstrained Binary Optimization model. Our analysis demonstrates that the dense and frustrated nature of the resultant landscape presents a formidable barrier to quantum optimization strategies, including the Quantum Approximate Optimization Algorithm, which typically struggle to locate global minima in glass-like systems \cite{farhi2014quantum}.
	
	To comprehensively validate the viability of this bio-computational paradigm, this study proceeds by first articulating the algorithmic architecture that transforms digital information into codon sequences and subsequently maps them onto secondary structures. We then rigorously evaluate the statistical quality of the generated ciphertext using the NIST SP 800-22 test suite to ensure indistinguishability from random noise. Following the statistical verification, we assess the quantum resistance of the primitive by simulating a Grover-based search and QAOA attacks on a coherent photonic quantum computer, quantifying the attack success probability against the dense QUBO formulation. Finally, we benchmark the computational throughput of the algorithm against industry standards such as AES and RSA to demonstrate its suitability for high-bandwidth applications \cite{sommerhalder2023hardware}. By reconciling the thermodynamic complexity of RNA folding with algorithmic efficiency, this work establishes the conceptual and operational validity of bio-inspired cryptography as a scalable defense mechanism for the post-quantum era.
	
	\begin{figure*}[htb!]
		\centering
		\includegraphics[width=\linewidth]{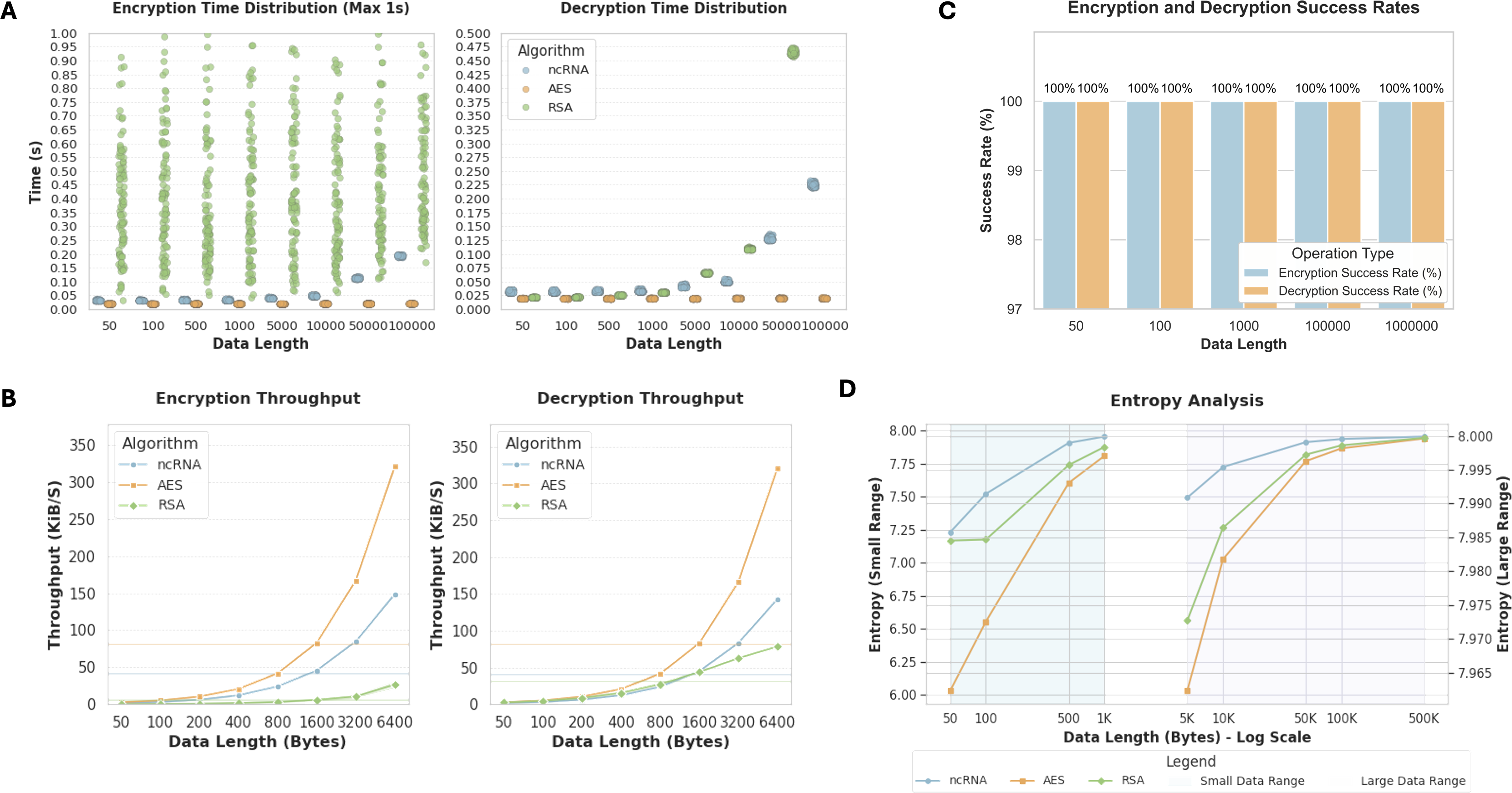}
		\caption{Summary of Algorithm Comparison and Testing Results. (A) Encryption/Decryption Efficiency. Comparison of average encryption and decryption times for Crypto-ncRNA (purple), AES-256 (green), and RSA-2048 (blue). Tests were performed on data lengths from 50 to 100,000 bytes. Time is plotted on a logarithmic scale. (B) Encryption/Decryption Throughput. Throughput in KiB/s is plotted against data length for Crypto-ncRNA, AES-256, and RSA-2048. (C) Ciphertext Randomness (Entropy). The average Shannon entropy per byte of the ciphertext generated by each algorithm is shown for various message lengths. The theoretical maximum entropy is 8.0 bits/byte. (D) Operational Reliability. The percentage of successful decryptions over 1,000 trials for each tested data length. All algorithms were tested on data sizes up to 1MB.}
		\label{fig:results_summary}
	\end{figure*}
	
	\section{Results}
\begin{figure*}[htb]
		\centering
		\includegraphics[width=\linewidth]{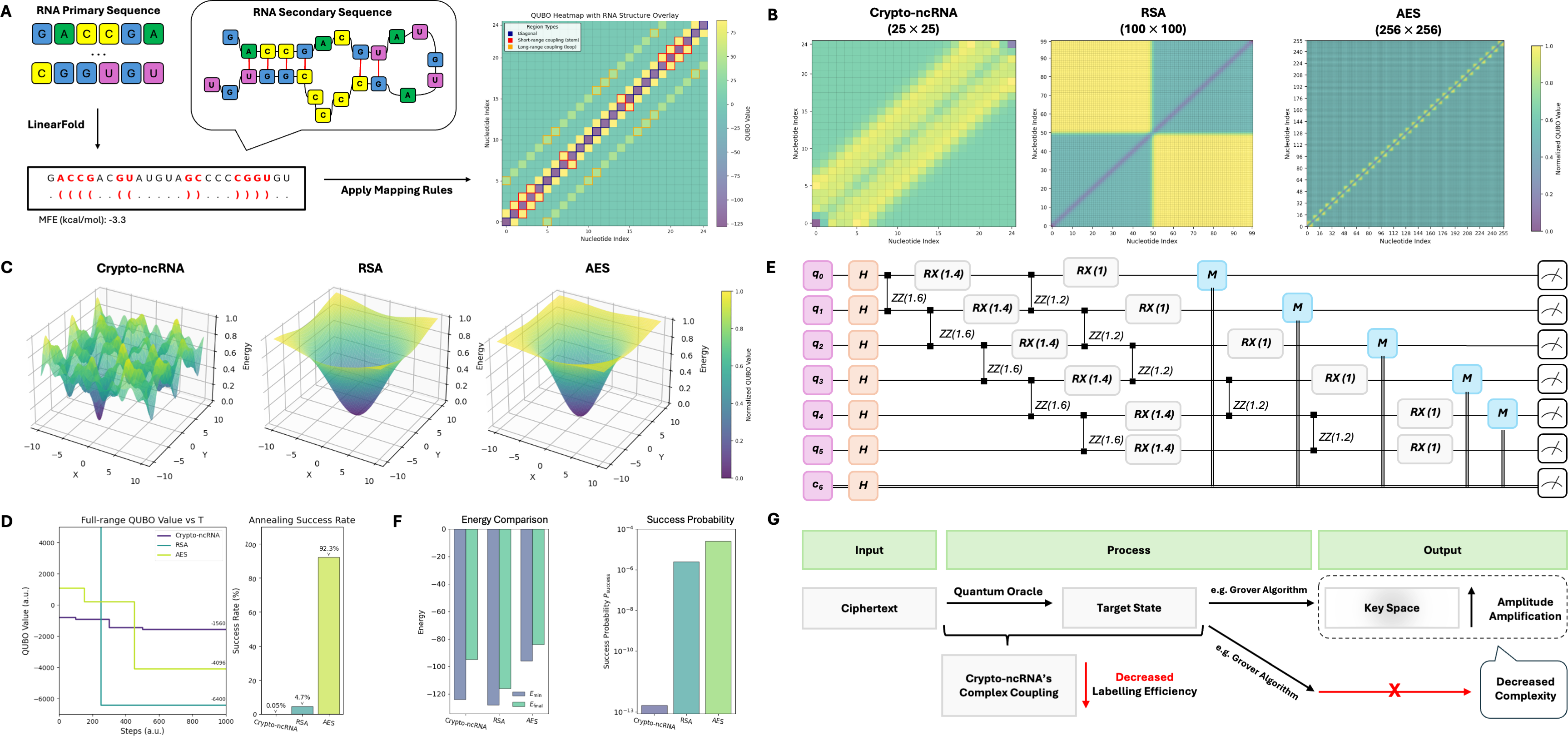}
		\caption{Summary of Quantum Test for Encryption Algorithm. (A) Mapping RNA Folding to QUBO. A diagram illustrating how RNA folding constraints, such as stem pairings and loop structures, are mapped to a banded QUBO matrix with short-range and long-range couplings. (B) QUBO Matrix Comparison. A visual comparison of the QUBO matrix structures for Crypto-ncRNA (banded dense), RSA (block-sparse), and AES (sparse diagonal). (C) Annealing Path Comparison. A conceptual depiction of quantum annealing paths for AES, RSA, and Crypto-ncRNA, showing the different energy landscapes they navigate. (D) Annealing Energy Evolution. A plot of energy evolution curves over annealing time for Crypto-ncRNA (purple), RSA (blue), and AES (green), based on 1000 samples. (E) QAOA Circuit Design. A schematic of the 6-qubit QAOA circuit used for the simulation, showing the initialization, Cost and Mixer layers, and mid-circuit measurement. (F) QAOA Success Probability. The calculated attack success probability for Crypto-ncRNA, RSA, and AES based on the final energy ($E_{\text{final}}$) achieved by the QAOA simulation relative to the theoretical minimum energy ($E_{\text{min}}$). A success is defined as the algorithm finding the exact ground state corresponding to the correct key ($E_{\text{final}} = E_{\text{min}}$). (G) Grover Attack Simulation. A workflow diagram illustrating how the complex coupling in Crypto-ncRNA's QUBO matrix affects the Oracle and amplitude amplification stages of a Grover-based quantum attack.}
		\label{fig:results_fig5}
	\end{figure*}

	\emph{Crypto-ncRNA} demonstrates consistent performance across heterogeneous computing environments. This has been validated through benchmarking against classical algorithms, namely RSA-2048 and AES-256 \cite{Rivest1978, Selent2010, FIPS197}. The subsequent sections and visualizations (Figures~\ref{fig:results_summary}, \ref{fig:results_fig5}, \ref{fig:framework}, and \ref{fig:principles_analysis}) detail its efficiency, throughput, ciphertext randomness, operational reliability, and the characteristics of the underlying key generation module. (Detailed metrics, methodologies, and raw data tables are presented in Appendix).
	
	\subsection{Biophysical complexity and structural avalanche effects}
\begin{figure*}[htb!]
		\centering
		\includegraphics[width=\linewidth]{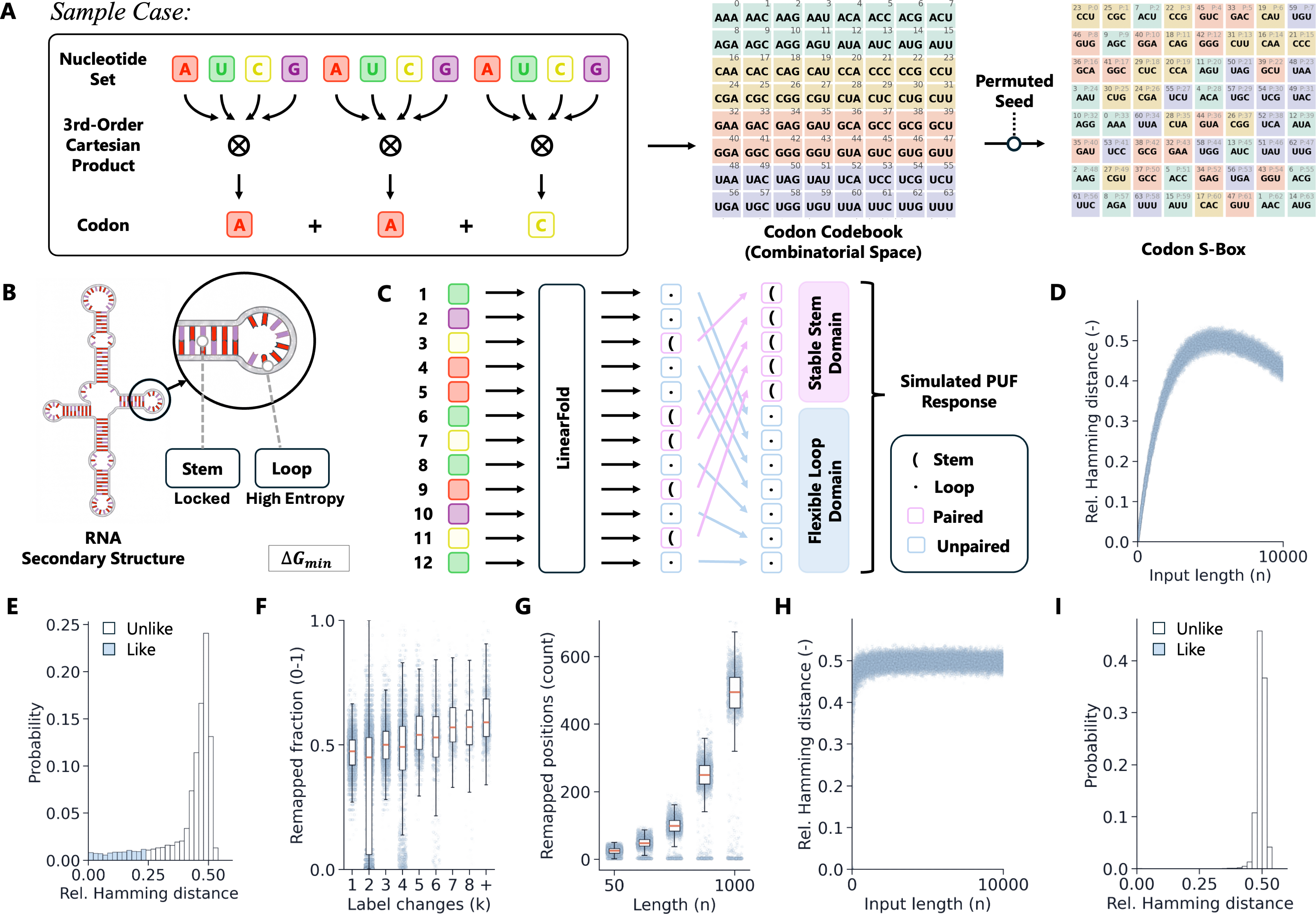} 
		\caption{Biophysical Mapping Principles and Statistical Validation of Structural Entanglement. 
			(A–C) Construction of Bio-inspired Primitives. The transformation from digital bits to biological complexity. (A) illustrates the Codon S-Box generation via a 3rd-order Cartesian product of nucleotides ($4^3=64$), mapping 6-bit inputs to a randomized codon codebook. (B–C) depict the Structural P-Box, where the LinearFold algorithm derives secondary structures (stems vs. loops) to topologically permute the sequence. 
			(D–G) Sensitivity and Error Intolerance Analysis. A statistical quantification of the algorithm's security properties. While (D–E) show the baseline randomness of the S-Box alone, (F–G) critically demonstrate the structural avalanche effect of the P-Box. The results indicate that even minimal errors in structural labeling (small $k$) trigger widespread index remapping. 
			(H–I) Cumulative Randomness Verification. Validation of the full system (S-Box + P-Box). The results show a rapid and stable convergence to an ideal Hamming distance of $\approx 0.5$ with over 99\% confidence.}
		\label{fig:principles_analysis}
	\end{figure*}

	To establish the biophysical foundation of our security model, we quantified the entropy introduced by the transformation from digital bits to biological complexity. As visualized in Figure~\ref{fig:principles_analysis}, the encryption pipeline integrates a baseline confusion layer, derived from a third-order Cartesian product of nucleotide bases, with a structural permutation layer that exploits the topology of RNA secondary structures. Our sensitivity analysis reveals that this architecture creates a critical error intolerance; minimal deviations in structural label prediction—simulating a corrupted key or approximate folding—trigger widespread index reordering. This phenomenon, effectively a biophysical avalanche effect, enforces a rigorous decryption condition where adversaries are prevented from utilizing approximate structures to derive partial permutation keys. Furthermore, cumulative randomness verification confirms that the synergistic application of codon mapping and structural permutation allows the ciphertext to converge rapidly to an ideal Hamming distance of approximately 0.5, yielding output that is statistically indistinguishable from random noise.
	
	\subsection{Statistical validation of ciphertext entropy}
	The statistical integrity of the generated ciphertext was rigorously verified against the full NIST SP 800-22 test suite (Table~\ref{tab:nist_results_main}), confirming that the biophysical encryption process yields cryptographic-quality output. Crucially, the system passed Maurer’s Universal Test (P-value = 0.85), a metric that evaluates the compressibility of the sequence. This result signifies that the output effectively behaves as a high-entropy source free from discernible patterns or algorithmic bias. By satisfying these stringent statistical standards, Crypto-ncRNA demonstrates that the chaotic nature of the simulated RNA folding landscape is successfully translated into a random bitstream capable of resisting statistical cryptanalysis.
	
	\begin{table}[h!]
		\centering
		\caption{Crypto-ncRNA’s NIST SP 800-22 Randomness Test Matrix Results}
		\label{tab:nist_results_main}
		\resizebox{\linewidth}{!}{%
			\begin{tabular}{llc}
				\toprule
				Test Name & P Value & Pass/Fail (P/F) \\
				\midrule
				Monobit Test & 0.5460386853638187 & P \\
				Frequency Within Block Test & 0.7963189024290873 & P \\
				Runs Test & 0.10786751132695774 & P \\
				Longest Run Ones in a Block Test & 0.22084938122535008 & P \\
				Binary Matrix Rank Test & 0.587708298333639 & P \\
				DFT Test & 0.8350238760410118 & P \\
				Non-overlapping Template Matching Test & 0.9999287413136844 & P \\
				Overlapping Template Matching Test & 0.43308028660774994 & P \\
				Maurer's Universal Test & 0.8514130113443941 & P \\
				Linear Complexity Test & 0.824328662851978 & P \\
				Serial Test & 0.507545477157905 & P \\
				Approximate Entropy Test & 0.5073170913186321 & P \\
				Cumulative Sums Test & 0.6611638690391457 & P \\
				Random Excursion Test & 0.1349606453103914 & P \\
				Random Excursion Variant Test & 0.039767475276814 & P \\
				\bottomrule
			\end{tabular}%
		}
	\end{table}
	
	\subsection{Computational throughput and operational efficiency}
	Beyond theoretical security, Crypto-ncRNA demonstrates operational efficiency suitable for high-bandwidth applications. Comparative benchmarking, as presented in Figure~\ref{fig:results_summary}, reveals that our method achieves orders-of-magnitude higher throughput than RSA-2048 and maintains a competitive performance profile relative to AES-256. Specifically, for data blocks of 100 KB, Crypto-ncRNA requires approximately 0.19s compared to 0.56s for RSA, positioning it as a viable candidate for replacing legacy systems in resource-constrained environments. This balance between computational complexity—required for security—and algorithmic efficiency confirms that the overhead of the RNA folding simulation is manageable for practical deployment without sacrificing resistance to quantum adversaries.
	
	\subsection{Resistance against quantum optimization and search algorithms}
	To empirically quantify the resilience of Crypto-ncRNA against quantum adversaries, we executed simulations on a coherent photonic quantum computer by mapping the RNA folding problem to a Quadratic Unconstrained Binary Optimization model. The experimental setup employed a six-qubit Quantum Approximate Optimization Algorithm circuit that incorporated initialization along with cost and mixer Hamiltonian layers. We introduced implicit hardware noise models typical of Noisy Intermediate-Scale Quantum devices, accounting for factors such as photon loss and coherence time limitations, while restricting the circuit depth to between one and four layers to mitigate severe decoherence. Under these conditions, the attack success probability was calculated based on the divergence between the final convergence energy and the theoretical minimum energy. The resulting probability of approximately $2.1 \times 10^{-13}$ confirms a robust theoretical defense. While this simulation validates the fundamental hardness of the energy landscape, we acknowledge the scale disparity between this six-qubit model and a full-scale attack, which would necessitate fault-tolerant quantum computing to process the complete matrix. Furthermore, the inherent ruggedness of the RNA folding landscape imposes a significant computational penalty on Grover-based search algorithms. Any oracle query within a Grover attack requires verifying the structural conformation, thereby forcing the adversary to incur an $O(N^3)$ computational cost for each superposition state evaluation and effectively neutralizing the quadratic speedup typically associated with quantum search.
	
	\section{Discussion}
	
	This study introduces Crypto-ncRNA as a post-quantum cryptographic primitive that leverages the simulated biophysical complexity of RNA folding to establish a security foundation orthogonal to traditional number-theoretic assumptions. While current global standardization efforts prioritize lattice-based and code-based cryptography \cite{chen2016report, alagic2022status, Bernstein2017}, these approaches predominantly rely on structured algebraic hardness. The inherent risk of such mathematical structures is the potential existence of undiscovered symmetries that could be exploited by novel algorithms, a vulnerability strikingly illustrated by the recent breakage of the SIKE algorithm (based on SIDH) via classical attacks targeting its auxiliary torsion points \cite{castryck2023efficient}. To mitigate the systemic risk of a monochromatic algorithmic landscape, it is imperative to explore security paradigms rooted in physical entropy rather than purely abstract mathematics. By exploiting the thermodynamic ruggedness of the RNA energy landscape—characterized by deep kinetic traps and frustration \cite{Tinoco1999, Chen2000, Solomatin2010}—Crypto-ncRNA provides a computational work-factor amplifier that is fundamentally distinct from the algebraic structures currently being standardized. In this framework, the industry-standard ChaCha20 cipher acts as the semantic security encapsulation layer, while the core quantum resistance is provided by the RNA-based key derivation process, ensuring that the system benefits from both established engineering reliability and novel bio-physical entropy.
	
	While the present work validates Crypto-ncRNA as a software algorithm, it functions conceptually as a ``digital twin'' for a future ncRNA-based Physical Unclonable Function (ncRNA-PUF). In this framework, the software implementation serves not merely as a simulation but as a high-fidelity verification of the thermodynamic information density inherent in RNA folding. By enforcing an algorithmic dependency on the secondary structure prediction, specifically utilizing the $O(N^3)$ complexity of the LinearFold algorithm \cite{huang2019linearfold, Zuker1981, Rivas1999}, we establish a computational barrier that mimics the physical intractability of measuring a specific molecular conformation without the correct environmental parameters. This approach aligns with the evolution of PUFs as essential trust anchors \cite{maes2010physically, maes2013physically, gao2020physical, Cambou2021}, extending the concept of biological feature extraction \cite{Zhou2021} to create a cryptographic primitive where the ``key'' is derived from the dynamic behavior of a virtual biomolecule. This establishes a theoretical security upper bound for the system prior to the introduction of wet-lab experimental constraints.
	
	Transitioning from this \textit{in silico} verification to a physical realization of an ncRNA-PUF presents engineering challenges centered on readout fidelity and noise management; however, recent biotechnological advances outline a concrete roadmap for implementation. The primary challenge of rapid and accurate readout is increasingly addressable via third-generation nanopore sequencing, where recent iterations such as the Oxford Nanopore R10.4 flow cell have demonstrated the capability to resolve homopolymers and single-nucleotide variations with precision sufficient for structural inference \cite{branton2008potential, deamer2016three, wang2021nanopore, ni2023benchmarking, zhang2023newest, sanderson2023comparison}. A more critical hurdle lies in the stochastic nature of biochemical reactions, which introduces insertion, deletion, and substitution errors that are typically fatal to cryptographic determinism. To bridge the gap between noisy biological substrates and precise digital keys \cite{Dodis2004Fuzzy}, the proposed architecture is compatible with advanced error-correction methodologies adapted from the DNA data storage domain. Specifically, the deployment of indel-correcting codes and robust decoding algorithms has proven effective in recovering error-free data from substantial sequencing noise \cite{Press2022, Hawkins2018, organick2018random, heinis2023survey, bencurova2023dna, Press2018, Press2020}. By integrating these high-fidelity readout technologies with rigorous error-correcting schemes, the thermodynamic entropy captured by our algorithm can be reliably extracted from physical substrates, suggesting that an ncRNA-PUF system is technologically feasible within the near-term horizon.
	
	The realization of such bio-physical security primitives offers distinct advantages over conventional silicon-based hardware security. Silicon PUFs, while widely deployed, have shown vulnerability to advanced machine learning modeling attacks and side-channel analysis, particularly when their challenge-response behaviors can be mathematically approximated \cite{Herder2014, gebali2022review, maiti2012systematic, becker2015gap, bhatta2024advancing, shepherd2021physical, kaur2021stratification}. Conversely, an ncRNA-PUF derives its security from the high-dimensional and non-linear interactions of molecular folding in solution, a process that resists electronic probing and fault injection attacks \cite{ruhrmair2013puf, wang2025deep}. This creates an orthogonal attack surface where security is anchored in chemical dynamics rather than semiconductor manufacturing variations \cite{Li2022, Luescher2024, Mondal2023, Park2024FourDimensional}. By integrating these biophysical properties with rigorous definitions of quantum unclonability \cite{Arapinis2021, zhang2019DNAOrigami}, Crypto-ncRNA represents a step toward ``bio-convergent security,'' positioning biological complexity not just as a medium for storage, but as an active computational shield for the post-quantum digital infrastructure.
	
	\begin{figure*}[htb!]
		\centering
		\includegraphics[width=\textwidth]{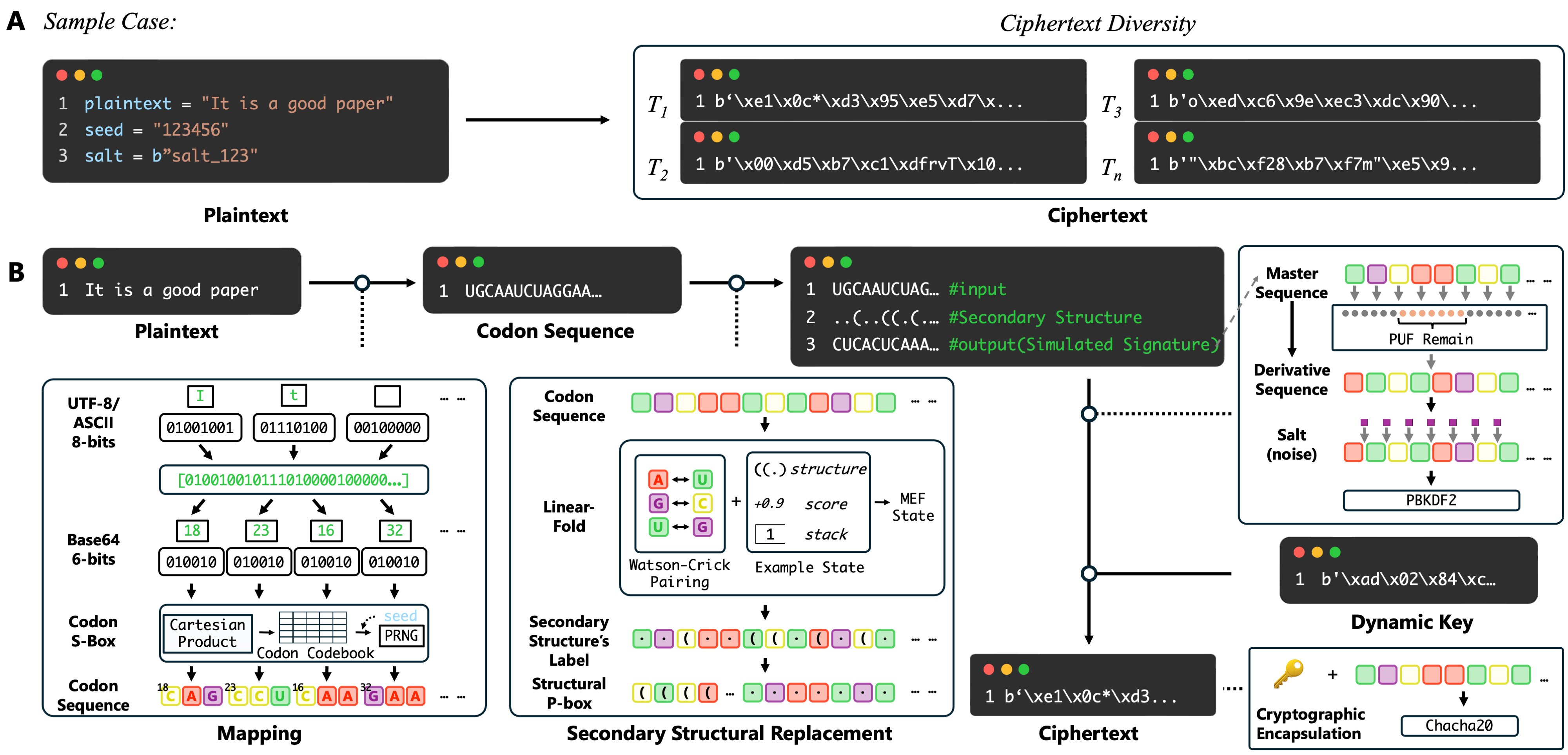}
		\caption{The Computational Workflow and Implementation Architecture of Crypto-ncRNA. 
			(A) Ciphertext Diversity. Demonstrates the algorithm's sensitivity to initialization parameters. A fixed plaintext combined with varying seeds and salts produces divergent ciphertexts ($T_1 \dots T_n$), ensuring high entropy. 
			(B) Encryption Pipeline. The architecture consists of four sequential stages: 
			1. Mapping: Plaintext is converted into a codon sequence via a dynamically generated Codon S-Box (derived from a Cartesian product of nucleotides). 
			2. Structural Replacement: The LinearFold algorithm calculates the Minimum Free Energy (MFE) secondary structure. This topological information drives a Structural P-box to permute the sequence based on stem-loop classifications. 
			3. Dynamic Key Generation: A Master Sequence is extracted from the folded structure to simulate a PUF response, which is then processed via PBKDF2 to derive a unique session key. 
			4. Encapsulation: The structurally entangled payload is finally encrypted using ChaCha20 with the derived dynamic key.}
		\label{fig:framework}
	\end{figure*}
	
	\section{Materials and Methods}
	The \emph{Crypto-ncRNA} framework establishes a bio-convergent cryptographic system by leveraging the biophysical properties of non-coding RNA (ncRNA). This paper presents a software algorithm, whose multi-tiered architecture (Figure~\ref{fig:framework}) transforms plaintext into secure ciphertext. While the algorithm is a self-contained software contribution, its security model is inspired by the long-term vision of a physical ncRNA-based Physical Unclonable Function (PUF). A physical PUF would provide resistance against cloning and physical attacks by introducing molecular-level variability, a concept that informs the design choices and security analysis of the software version presented here.
	
	\subsection{The Crypto-ncRNA Computational Framework}
	The Crypto-ncRNA algorithm functions as a \textit{biophysically-inspired, dual-layer cryptographic system}. Operating entirely \textit{in silico}, it translates the thermodynamic complexity of RNA folding into a computational hardness assumption.
	
	\subsubsection{Data Encoding and Cartesian S-Box Generation}
	The transformation from the binary to the biological domain is achieved through a third-order Cartesian product of the nucleotide set $\{A, U, C, G\}$, which establishes a complete codon codebook comprising 64 unique combinations corresponding to a 6-bit binary input space. Defined as $\mathcal{S}_{\text{codon}} = \{A, U, C, G\}^3 \rightarrow \{0, \dots, 63\}$, this mapping provides the substrate for information encoding. To introduce dynamic non-linearity, a user-defined seed drives a Pseudo-Random Number Generator (PRNG) to shuffle the codebook indices, thereby creating a randomized Codon S-Box. This initial substitution layer ensures that identical plaintext inputs map to divergent codon sequences under varying initialization vectors, effectively preconditioning the data before the biological folding simulation.
	
	\subsubsection{Security Architecture and Complexity Amplification}
	The algorithm integrates structural dependencies with statistical masking by utilizing the Structural P-Box, a primitive that exploits the computational intractability of RNA secondary structure prediction. Specifically, the LinearFold algorithm is employed to calculate the Minimum Free Energy (MFE) state, yielding a dot-bracket notation that classifies nucleotides into paired stems or unpaired loops. This topological information rigorously dictates the permutation rules, where the sequence is reordered based on structural classifications according to $\mathbf{S}_{\text{folded}} \leftarrow \text{FoldPermute}(\mathbf{S}_{\text{RNA}}, \text{Structure})$. By linking the permutation logic directly to the thermodynamic state, the system enforces a high sensitivity to input errors; minor deviations in the predicted structure trigger substantial disruptions in the final sequence order. Consequently, inverting this permutation without the precise structural key necessitates solving the RNA inverse folding problem, thereby amplifying the computational work-factor.
	
	To secure this structural information, the folded sequence and permutation indices are encapsulated using the ChaCha20 symmetric stream cipher \cite{bernstein2008chacha}:
	\[ 
	\mathcal{C} = \text{ChaCha20}\left(\mathbf{S}_{\text{folded}} \parallel \mathbf{I}_{\text{perm}},\; \mathbf{K}_{\text{dynamic}} \right) \parallel \text{Hash}(\mathbf{S}_{\text{folded}})
	\]
	This architecture serves as a computational work-factor amplifier for the pre-shared secret. In scenarios involving brute-force or Grover's algorithm attacks \cite{grover1996fast}, verifying a candidate key $\mathbf{K}_{\text{dynamic}}$ requires replicating the folding of the specific pre-shared seed (or password). Because the key derivation is algorithmically bound to the thermodynamic state, validation necessitates running the computationally intensive RNA folding simulation (LinearFold, $O(N^3)$) to derive the correct $\mathbf{K}_{\text{dynamic}}$ from any guessed seed. This requirement imposes a significant computational penalty on every oracle query performed by an adversary trying to recover the seed.
	
	\subsection{Simulated Biophysical Key Derivation}
	While inspired by the theoretical concept of an ncRNA-based Physical Unclonable Function (PUF), the current implementation operates as a \textit{Simulated PUF} within the software domain.
	
	\subsubsection{Dynamic Key Generation from Structural Topology}
	Rather than relying on simple static credentials, the encryption key $\mathbf{K}_{\text{dynamic}}$ is derived dynamically from the unique topological signature of an RNA sequence generated from a pre-shared secret (or password)~\cite{Zhou2021}. A subsequence of this secret-derived folded structure is extracted to serve as a virtual biological response, which is subsequently processed through a key stretching algorithm defined by $\mathbf{K}_{\text{dynamic}} = \text{PBKDF2-HMAC-SHA256}\left(\Phi(\mathbf{S}_{\text{secret}}), \text{Salt}, 10^4 \right)$ \cite{Kaliski2000}.
	
	\subsection{Mapping RNA Folding Constraints to QUBO}
	To quantify quantum hardness, the RNA folding problem is formulated as a Quadratic Unconstrained Binary Optimization (QUBO) model \cite{glover2018tutorial}. For an RNA sequence of length $N$, the state of each nucleotide position $i$ is represented by a binary variable $x_i \in \{0,1\}$, where $x_i=1$ denotes the formation of a base pair. The thermodynamic landscape is governed by the energy function $H_{\text{RNA}} = \sum_{i<j} J_{ij}x_ix_j + \sum_i h_i x_i$, where $J_{ij}$ corresponds to pairing energies consistent with Watson-Crick rules. This Hamiltonian is subsequently transformed into the standard QUBO matrix form $H_{\text{QUBO}} = \mathbf{x}^T Q \mathbf{x}$. The resulting matrix $Q$ incorporates dense coupling terms ($Q_{i,i+1} \approx 88$) derived from stem stacking constraints and long-range couplings ($Q_{i,i+5} \approx 44$) introduced by loop permutation rules, creating cross-dimensional interference that generates a rugged energy landscape resistant to quantum optimization \cite{kadowaki1998quantum}.
	 
	The matrix $Q$ incorporates dense coupling terms ($Q_{i,i+1} \approx 88$) derived from stem stacking constraints, and long-range couplings ($Q_{i,i+5} \approx 44$) introduced by the loop permutation rules. These constraints create cross-dimensional interference, resulting in a rugged energy landscape difficult for quantum optimizers to traverse.
	
	\subsection{Quantum Resistance Verification via QAOA}
	We evaluated the algorithm's resilience using the Quantum Approximate Optimization Algorithm (QAOA). The target is to find the ground state of the Hamiltonian $H_C$ encoding the QUBO problem:
	\[ H_C = \sum_{i} Q_{ii} Z_i + \sum_{i<j} Q_{ij} Z_i \otimes Z_j \]
	The QAOA evolution involves alternating applications of the cost Hamiltonian $H_C$ and a mixer Hamiltonian $H_M = \sum X_i$:
	\[ |\psi(\gamma, \beta)\rangle = \prod_{k=1}^p e^{-i\beta_k H_M} e^{-i\gamma_k H_C} |+\rangle^{\otimes n} \]
	The success probability of a quantum attack is defined as the probability of measuring the state corresponding to the correct key (global minimum energy $E_{\min}$):
	\[ P_{\text{success}} = |\langle \mathbf{x}_{\text{key}} | \psi(\gamma, \beta) \rangle|^2 \approx \exp\left(-(E_{\text{final}} - E_{\min})\right) \]
	Our simulations on a coherent photonic quantum computer (CPQC-1) \cite{QBosonKaiwuSDK} indicate that Crypto-ncRNA's QUBO matrix exhibits a negligible success probability ($P_{\text{success}} \approx 2.1 \times 10^{-13}$), confirming robust resistance against near-term quantum attacks.
	
	\subsection{Implementation, Benchmarking, and Statistical Validation}
	The Crypto-ncRNA framework was implemented in Python (v3.12), utilizing standard libraries for numerical computation and cryptographic operations to ensure reproducibility. Comparative benchmarking was conducted against classical algorithms using the pycryptodome library (v3.21.0) \cite{pycryptodome}, specifically employing AES-256 in Cipher Block Chaining (CBC) mode with PKCS7 padding \cite{kaliski1998pkcs7} and RSA-2048 with PKCS\#1 Optimal Asymmetric Encryption Padding (OAEP) \cite{moriarty2016pkcs1}. To verify the cryptographic integrity of the generated ciphertext, the output was subjected to the NIST SP 800-22 Statistical Test Suite \cite{rukhin2001statistical, bassham2010sp}. This assessment involved analyzing ciphertext samples from multiple independent trials across varying message lengths to rigorously quantify indistinguishability from random noise, ensuring the system meets established security standards for cryptographic applications.
	
	\section*{Acknowledgments}
	This work was supported by the Natural Science Foundation of Xinjiang Uygur Autonomous Region (Grant Number: 2024D01C216) and the “Tianchi Talents” introduction plan. We acknowledge QBoson Quantum Technology for providing access to their CPQC-1 quantum computer and Kaiwu SDK for the quantum evaluation portion of this study. 
	
	\section*{Declarations}
	\subsection*{Author Contributions}
	X.W., Y.W., and T.-Y.H. designed and performed research; T.-Y.H. prepared figures; K.W. provided funding, supervision, and contributed to manuscript editing; and X.W., Y.W., T.-Y.H., and Z.J. contributed to manuscript editing.
	
	\subsection*{Competing Interests}
	The authors declare no competing interests.
	
	\subsection*{Code Availability}
	The code used in this article can be obtained from GitHub: \url{https://github.com/JLU-WangXu/crypto-ncRNA}
	
	\bibliography{reference}
	\bibliographystyle{unsrt}
	
\end{document}